\documentclass[fleqn,usenatbib]{mnras}

\usepackage{newtxtext,newtxmath}

\usepackage[T1]{fontenc}

\DeclareRobustCommand{\VAN}[3]{#2}
\let\VANthebibliography\thebibliography
\def\thebibliography{\DeclareRobustCommand{\VAN}[3]{##3}\VANthebibliography}

\usepackage{graphicx}	
\graphicspath{{gfx/}}
\usepackage{adjustbox}
\usepackage{amsmath}	
\usepackage{gensymb}
\usepackage{subcaption}
\usepackage{listings}
\usepackage{longtable}

\usepackage{xspace}
\usepackage{pgf}
\newcommand{\gaia}{\textsl{Gaia}\xspace}

\newcommand{\nuna}[2]{(#1)\,\textit{#2}} 

\PassOptionsToPackage{dvipsnames}{xcolor} 
\usepackage{pstricks}
\usepackage{color}
\usepackage{ulem}
\usepackage{color,soul}
\usepackage{textcomp}

\title[\gaia DR3 spectra of V-types]{Spectral analysis of basaltic asteroids observed by the \gaia space mission}

\author[D Oszkiewicz et~al.]{
Dagmara Oszkiewicz$^{1}$\thanks{orcid=0000-0002-5356-6433} ,
Hanna Klimczak$^{1}$ ,
Benoit Carry$^{3}$, 
Antti Penttil\"{a}$^{2}$, \and
Marcel Popescu$^{4}$,
Joahim Kruger$^{1}$,
Marcelo Aron Keniger$^{4,5}$
\\
$^{1}${Astronomical Observatory Institute, Faculty of Physics, Adam Mickiewicz University, S{\l}oneczna 36, 60-286 Pozna{\'n}, Poland}\\
$^{2}${Department of Physics, P.O. Box 64, FI-00014 University of Helsinki, Finland}\\
$^{3}${Universit\'e C{\^o}te d'Azur, Observatoire de la C{\^o}te d'Azur, CNRS, Laboratoire Lagrange, France} \\
$^{4}${Astronomical Institute of the Romanian Academy, 5 Cutitul de Argint, 040557, Bucharest, Romania}\\
$^{4}${Nordic Optical Telescope, Rambla Jos\'e Ana Fern\'andez P\'erez 7, E-38711 
Breña Baja, Spain}\\
$^{5}${Stellar Astrophysics Centre, Department of Physics and Astronomy, Aarhus 
University, Ny Munkegade 120, DK-8000 Aarhus C, Denmark
}
}

\date{Accepted XXX. Received YYY; in original form ZZZ}

\pubyear{2022}

\begin{document}
\label{firstpage}
\pagerange{\pageref{firstpage}--\pageref{lastpage}}
\maketitle

\begin{abstract}
There is a great deal of scientific interest in characterizing the basaltic asteroids (spectrally classified as V-types), as they are the key to understanding planetesimal formation and evolution in the early Solar System. These have long been recognized as parts of the crusts of fully differentiated planetesimals. Thus, their multiplicity, distribution, and physical characteristics are crucial for providing context for and constraining the theoretical evolution models of the Solar System. In this work, we perform spectral analysis with an extended data set of spectral measurements from the ESA \gaia mission Data Release 3, thus increasing the sample size of the analyzed V-types by more than three times as compared to the literature. Using the data provided by \gaia we identified $\sim$2000 possible V-type asteroids. About 350 of them successfully pass our data validation criteria. This sample includes 31 new V-type asteroids beyond 2.5\,au and 6 in the Phocaea region. We confirm that the V-type asteroids in the middle and outer part of the main belt show distinct spectral properties compared to typical vestoids. In the inner main belt, we found a great diversity of spectral parameters among the V-types in all populations. Number of asteroids show band depths even greater than that of (1459) Magnya. Furthermore, some objects present 0.9~\textmu{}m band-centers more than one standard deviation away from the typical value for vestoids. However since the DR3 band centers are often overestimated, those findings are to be confirmed. Overall our results indicate that the inner main belt may contain remnants of multiple differentiated planetesimals, not just (4) Vesta. 

\end{abstract}




\section{Introduction}

Planetesimals were the first large solid bodies that formed in the Solar System more than 4.5 billion years ago. Current Solar System evolution theories and abundant meteoritic evidence suggest that there once existed 30 to 150 different differentiated planetesimals (into an iron core, silicate mantle, and crust) \citep{burbine2002meteoritic, scott2009oxygen, scott2015early, greenwood2015geochemistry}. They were then collisionally disrupted during later stages of the Solar System evolution. Studies of various isotopes in iron meteorites confirm that these extensive collisions exposed iron cores to cool in a time frame of 7.8 to 11.7 million years after the formation of calcium-aluminium-rich inclusions \citep{hunt2022dissipation}. Numerical studies constrain the formation location of differentiated planetesimals to the present terrestrial planet region \citep{bottke2006iron} or to a wider formation ranges at orbital locations between 1.3~au and 7.5~au \citep{lichtenberg2021bifurcation}. The disrupted fragments of planetesimals can be found in the current Main Asteroid Belt and in meteorite collections today. In particular, V-type asteroids are identified as parts of the crusts of thermally evolved planetesimals.

V-type asteroids are immediately recognizable from the spectra by two prominent wide olivine/pyroxene absorption bands at wavelengths around 1.0~\textmu{}m and 2.0~\textmu{}m. The first identified V-type object was the asteroid (4) Vesta, which is the largest object of this taxonomic class and is considered a fossil planetesimal. Due to characteristic spectra, Vesta was immediately linked to howardite-eucrite-diagonite meteorites (HEDs) in the early 1970s \citep{mccord1970asteroid}. A few decades later, new V-types started to be found in the vicinity of (4) Vesta and in the near-Earth asteroid population \citep{binzel1993chips, burbine2001vesta, Florczak2002Discovering}. Soon, the V-types close to Vesta were recognized as Vesta family \citep{williams1989asteroid, nesvorny2015identification, hirayama1918groups}. This was reinforced by the discovery of two large impact craters in Vesta \citep{thomas1997impact, mcsween2013dawn, mcsween2010hed, marchi2012violent, schenk2012geologically}. The link between most HED meteorites, V-type asteroids, and (4) Vesta is one of the best established in planetary science.

Another milestone for studies of V-type asteroids was the discovery of the basaltic nature of asteroid (1459) Magnya in early 2000 \citep{lazzaro2000discovery}. Based on dynamical investigation \cite{michtchenko2002origin} suggested that this $\sim$17~km \citep{delbo2006midi} asteroid located beyond 2.8~au from the Sun is most probably a part of other (than Vesta) planetesimal that existed in the outer parts of the main belt. \cite{hardersen2004mineralogy} further substantiated this claim by finding the discordant pyroxene chemistry of Magnya compared to that of Vesta. This showed that indeed there might be fragments of differentiated planetesimals other than Vesta in the Solar System supporting the hypothesis of multiple differentiated planetesimals.

More discoveries of such objects were made possible because of the start of a big survey era. Multiple studies focused on the search for V-type asteroids beyond 2.5~au in the middle and outer main belt \citep{roig2006selecting, roig2008v, moskovitz2008spectroscopically, duffard2009two, solontoi2012avast, oszkiewicz2014selecting, licandro2017v, medeiros2019compositional, mansour2020distribution}. About a dozen new V-types were found. Spectral studies showed that these objects similarly to Magnya are distinct from Vesta \citep{ieva2016spectral, ieva2018basaltic, leith2017compositional}. Furthermore, \cite{roig2008v} showed that large asteroids ($>$ 5~km) in the middle main belt region have a low probability of about $\sim$1~\% of having evolved from Vesta through a combination of the Yarkovsky effect and dynamical resonances. \cite{ieva2018basaltic} suggest that the Eos asteroid family in the outer main belt could be the source of some of the V-types in this region. Dynamical evolution of V-type candidates also indicates that the parent bodies of the Eunomia and Merxia / Agnia families could also be a plausible source of V-types in the mid and outer main belt \citep{carruba2014dynamical}. Similar origins are also suggested by \cite{migliorini2021characterization}. On the other hand, \cite{brasil2017scattering} suggest the 'jumping Jupiter' planetary migration scenario as a possible transport mechanism from Vesta.

Theoretically, the number of V-type fragments originating from different planetesimals should be even greater in the inner Main belt \citep{bottke2006iron}. However, most of the V-type asteroids in the inner main belt are part of the Vesta family or are considered fugitives, i.e., objects that evolved away from the family and are now beyond recognition as family members with traditional clustering methods. Early studies showed that some of the V-types in the inner main belt present a deeper 1.0~\textmu{}m absorption bands than Vesta \citep{Florczak2002Discovering}. However, these findings were not confirmed by \cite{ieva2016spectral, ieva2018basaltic}. Only one asteroid in the inner main belt, (2579) Spartacus, shows an unusually low band area ratio (BAR, ratio of the areas encompassed by the 1.0- and 2.0-\textmu{}m absorption bands) compared to Vesta family members \citep{burbine2001vesta, moskovitz2010spectroscopic, oszkiewicz2019physical}. However, backward dynamical integration shows that the object could well originate from the Vesta family. Furthermore, BAR is often considered an unreliable parameter.

Basaltic material unrelated to Vesta is found in so-called anomalous meteorites \citep{zhang2019oxygen, yamaguchi2006anomalous, barrett2017mineralogy, mcsween2010hed} and micrometeorites \citep{gounelle2009unique}. One of those falls was observed and its trajectory was traced back to the inner main belt \citep{spurny2012bunburra, bland2009anomalous}.

An unexpected place to find basaltic material was on the surface of carbonaceous asteroid (101955) Bennu. \cite{dellagiustina2021,tatsumimnras21} reported,  based on data provided by OSIRIS-REx space mission, the identification of several bright boulders  which showed the absorption bands characteristic of pyroxene composition, and their spectral features were matched by HED composition. At first glance there is a dynamical similarity between the inner main belt basaltic asteroids and the source regions of Bennu \citep{tatsumimnras21} which can explain how these boulders were implanted on Bennu. These findings provide new constraints on the collisional and dynamical evolution of the inner main belt.

\cite{oszkiewicz2015differentiation, oszkiewicz2017non} showed that some V-types in the inner main belt have spin orientation and thus Yarkovsky drift direction that is inconsistent with the origin in the Vesta family. However, strong evidence is still lacking, and the question of possible non-vestoids in the inner main belt is thus still under debate.

In this work, we identify V-type asteroids across the Solar System from \gaia  Data Release 3 (DR3) spectral data. We analyze their spectral parameters in the context of multiple differentiated planetesimals. In Section \ref{data} we describe the data used in the analysis and in Section \ref{met} the methods used for spectral type classification and spectral analysis. Results are presented in Section \ref{res}, discussion and conclusions are in Sections \ref{disc} and \ref{conl}.

\section{Data}
\label{data}
\subsection{Gaia DR3 data}
European Space Agency's \gaia mission was launched in December 2013. It operates at the Sun-Earth Lagrangian L2 point with the main goal of creating a 3-dimensional map of our Galaxy by surveying more than a billion stars
\citep{prusti2016gaia}. \gaia measures the positions, proper motions, brightness and spectra of objects crossing its fields of view, including asteroids \citep{2007EMP..101...97M, tanga2012solar, tanga2022data}. 

The current third data release (DR3) includes, for the first time, the mean reflectance spectra of 60,518 asteroids at visible wavelengths (0.374~\textmu{}m -- 1.05~\textmu{}m), which is a 10-fold increase in the number of known asteroid spectra \citep{gaiaspec}. This huge increase in the number of asteroid spectra is expected to result in a new taxonomic scheme \citep{gayon2012asteroid} and more detailed understanding of the distribution of various taxonomies across the Solar System and among asteroid families. 

The \gaia spectral data are obtained by the blue and red photometers (BP/RP), which also cover the UV spectral region and are collectively referred to as XP instruments \citep{prusti2016gaia, brown2016}. Details of observations and data processing for asteroids are described in \cite{gaiaspec}. For each object, average spectra sampled at 16 wavelengths (in the range between 0.374~\textmu{}m and 1.034~\textmu{}m every 0.044~\textmu{}m) together with 16 quality flags are provided. These spectra are the result of averaging typically more than 20 individual epoch spectra for each object \citep{gaiaspec}. One epoch corresponds to a single transit of an asteroid in the \gaia focal plane.

\subsection{Training data set}
\begin{figure}
    \centering
    \input{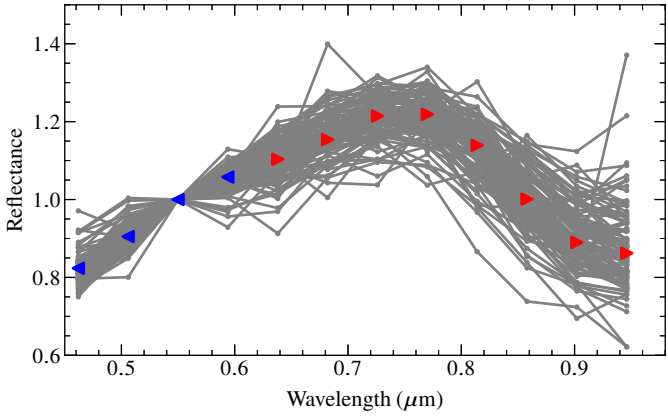}
    \caption{DR3 Spectra of 149 V-type asteroids used in the training process and their averaged spectra from the BP (blue) and RP (red) spectrometers. We marked average reflectances from BP and RP spectrometers in blue and red, respectively.
    }
    \label{vtypes_gaia}
\end{figure}

We classify asteroids into Bus-DeMeo classes \citep{demeo2009extension} using three different supervised methods (see Section \ref{met}), which require a training data set. For that purpose, we used the collection of 3057 taxonomic labels from ground-based visible spectra from \citet{mahlke2022}. Those labels where then cross-matched by asteroid number or designation with objects observed by the \gaia mission leading to a sample containing 149 V-types and 2908 other-type asteroids (Fig. \ref{vtypes_gaia}).

By using the \gaia data itself as a training set we teach the algorithms to recognize any possible systematic biases in the data that might be intrinsic to the \gaia mission. For example, the ground-based data are often obtained at opposition, thus at small phase angles. \gaia measurements are taken at phase angle $>$10 degrees, this might result in moderate change of spectral slopes due to phase-reddening effect
\citep{sanchez2012phase,BINZEL201941}.

In the training parameter set we omit reflectances at the wavelengths of 0.374, 0.418, 0.990, and 1.034~\textmu{}m, as they are known to be affected by large random and systematic errors \citep{gaiaspec}. We perform a linear regression using the remaining 12 points to obtain the spectral slope. This slope is removed from the spectra and treated as one of the parameters. Slope and slope-removed reflectances at the 12 wavelengths are then the final set of parameters for each spectra used in the training process. In Fig.~\ref{vtypes_gaia} we show the V-type spectra used in training.

We also note that prior to using the \gaia DR3 data itself in the training process we experimented with several different training data sets from ground-based observations to simulated clones. The highest accuracy was reached when we used the DR3 spectra in the training process. We also inspected the predicted V-types visually to confirm the correctness of the results. We have assessed the influence of different parameters used in the classification task on the resulting accuracy as well. Using slope and reflectances vs. only reflectances from raw (no slope removed) spectra resulted in a similar accuracy.

\section{Methods}
\label{met}

\subsection{Machine learning methods}
We perform a binary classification into V-type vs. other-types categories to increase the success rate. We use the three best-performing methods from \cite{klimczak2021predicting, klimczak2022}. We briefly summarize the methods below; for more details, see our recent works \cite{klimczak2021predicting, klimczak2022}. 

\begin{itemize}
\item {\bf Gradient Boosting (GB)} is a set of weak prediction methods. In particular, we use several decision trees. In each iteration, a new model is trained to minimize the residual left from the estimator from the previous stage \citep{elements-of-statistical-learning}.
\item {\bf Support Vector Machines (SVM)} divide the feature space of different classes into separate regions with hyperplanes \citep{elements-of-statistical-learning}. The algorithm maximises the width of the gap between the different classes.
\item {\bf Multilayer Perceptrons (MLP)} we use a feedforward artificial neural network with two or three layers of perceptrons, each with 32 or 64 neurons. The weights of the neural connections are adjusted in a learning process with learning rates between 0.01 and 0.1 \citep{GoodBengCour16}. 

\end{itemize}
We employ the implementation of methods from the Python scikit-learn package \citep{pedregosa2011scikit}. To evaluate the algorithms, we use a balanced accuracy metric:

\begin{align*}
    \text{BAcc} &= \frac{1}{k} \sum_{i=1}^{k}{\frac{TP_i}{TP_i + FN_i}} 
\end{align*}

where $k$ is the number of classes, $TP_i$ (``true positives'') is the number of correctly classified objects of class $i$, $FP_i$ (``false positives'') is the number of objects incorrectly classified in class $i$ and $FN_i$ (``false negatives'') the number of incorrectly classified objects from class $i$. This metric gave the same general conclusions as unbalanced accuracy, F1 score, and Matthews correlation coefficient in our previous work \citep{klimczak2021predicting} and is known to work well with imbalanced problems, as is the case in our study \citep{Kelleher_etal2015}. To estimate the accuracy of the method, we use a 5-fold validation. That is, we divide the training set into 5 random parts. The algorithms are trained in 4 parts and evaluated in the 5th part. This process is repeated five times.

For the final prediction on new asteroids, the models are trained on the whole dataset of 3057 objects, 149 of which are V-types.

\subsection{Spectral parameters}
\label{spectral_params}
To analyze the \gaia spectra of classified V-type asteroids in a consistent way with the literature \citep[e.g.,][]{ieva2016spectral, matlovivc2020spectral} we compute parameters that characterise the shape of the 0.9~\textmu{}m absorption band on the original reflectances. In particular, for each spectrum we determine the reflectance gradient in the 0.5~\textmu{}m--0.75~\textmu{}m range (slope A), the reflectance gradient in the 0.8~\textmu{}m--0.92~\textmu{}m (slope B), apparent depth and center of the 0.9~\textmu{}m band (B1). Steeper A-slopes are an indication of space weathering \citep{fulvio2012space, fulvio2016spectral} and a higher band depth relates to a larger grain size, the presence of fresh/unweathered pyroxene, or a different mineralogy \citep{cloutis2013spectral}. Gradients are computed as the slopes of the linear regression performed in the given wavelength ranges.

\begin{figure}
    \centering
    \includegraphics[width=0.5 \textwidth]{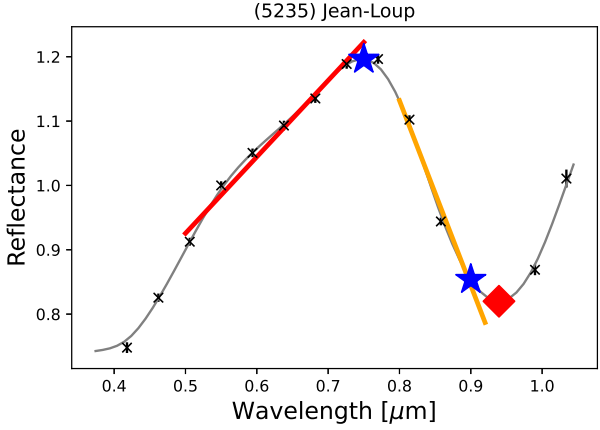}
    \caption{\gaia spectrum of (5235) Jean-Loup (black) and linear regression fits for wavelength ranges 0.5~\textmu{}m--0.75~\textmu{}m (red, slope A) and 0.8~\textmu{}m--0.92~\textmu{}m (orange, slope B). Grey line represents fitted cubic spline, reflectances at wavelengths used in computation of band depth are denoted with stars. Center of the 0.9~\textmu{}m absorption band is denoted with a diamond symbol. DR3 uncertainties are on the order of marker size.}
    \label{5235}
\end{figure}

To estimate the band depth, we fit cubic splines to all \gaia points and then compute the reflectance at 0.75~\textmu{}m and 0.9~\textmu{}m (apparent depth) and their ratio. Minimum of the 0.9~\textmu{}m band is found as the minimum of the fitted smoothing spline. In Fig. \ref{5235} we illustrate the parameters for the spectrum of a V-type asteroid, (5235) Jean-Loup. The derived parameters for this objects are slope A=11.9~\%/10$^3$~\AA, slope B=-28.7~\%/10$^3$~\AA, band depth = 1.41.

\begin{figure}
    \centering
    \input{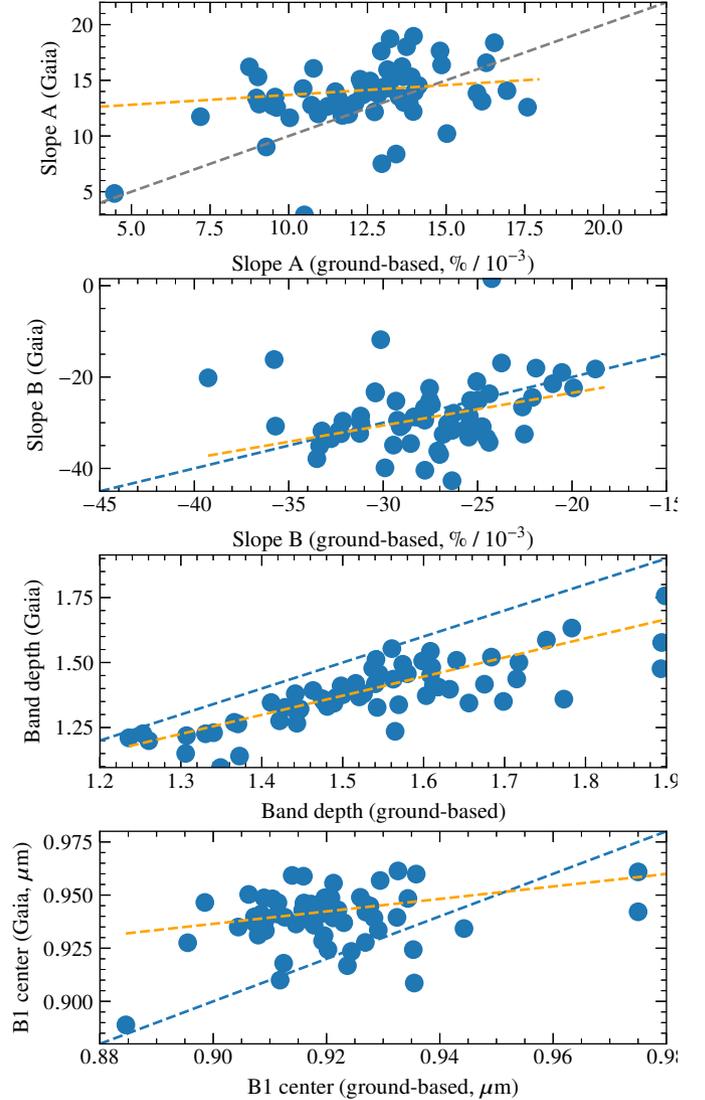}
    \caption{Comparison of spectral slopes,
    band depths and centers for ground-based and \gaia observations of V-type asteroids.
    The dashed blue and orange lines represent the 1:1 relation,
    and a linear regression on the sample, respectively.}
     \label{comp}
\end{figure}

\cite{gaiaspec} found that \gaia slightly underestimates the depth of the 0.9~\textmu{}m absorption band and the spectral slope of V-types compared to the SMASSII survey \citep{bus2002phase}. Furthermore, a systematic reddening of the UV part of the DR3 spectra as compared to the ground-based observations was found by Tinaut-Ruano et al., in prep. This reddening was contributed to the choice of solar analogue used to reduce the DR3 spectra (Tinaut-Ruano et al., in prep.). At the red part of the spectrum the choice of solar analogue was found not to affect the asteroid spectra. However differences wrt. ground-based observations in this part of the spectrum are also known to exist. To verify this we compare spectral slopes, band depths, and band centers estimated from ground-based observations from \cite{mahlke2022} with those obtained by the \gaia mission. In Fig.~\ref{comp} we present the spectral parameters (slope A, slope B, 0.9~\textmu{}m band depth and center) for V-types present in both data sets. Distribution of slope A and slope B correlate with the ground-based data and there is no significant systematic bias. However, the DR3 slopes seems to be overestimated for objects with small ground-based slope values and underestimated for objects with larger ground-based slope values. Band depth, on the other hand, is indeed systematically underestimated for all asteroids by about 0.15. Band centers seem to be overestimated for most objects. The linear relations between the DR3 and ground-based spectra are:

\begin{eqnarray*}
\text{slope}_\text{A}^\text{DR3} &=& 0.323 \times \text{slope}_\text{A}^\text{grb} + 9.975\\
\text{slope}_\text{B}^\text{DR3} &=& 0.713 \times \text{slope}_\text{B}^\text{grb} -9.207 \\
\text{apparent depth}^\text{DR3} &=& 0.679 \times \text{apparent depth}^\text{grb} + 0.349 \\
\text{BI center}^\text{DR3} &=& 0.188 \times \text{BI center}^\text{grb} + 0.767,
\end{eqnarray*}

where the upper indices denote the Gaia (DR3) and ground-based (grb) parameters. To determine the linear fits we have used the RANSACRegressor from the sklearn package with 3-sigma clipping. Generally, the spectra are internally consistent and thus can be internally compared.

\subsection{Validation of V-type candidates}

We follow a two-level validation. For the selection of V-type candidates we require that the DR3 quality flags for all used wavelengths are validated. Additionally all the predicted V-types from the full sample in the middle main belt, outer main belt, and the Phoacea and Hungaria regions are validated visually and added to the validated sample.

For the analysis of spectral parameters we require that the average signal-to-noise snr $>$ 40 and the DR3 spectral quality flags are validated for the wavelengths used in parameter estimation.

\section{Results}

\label{res}

Based on the training data set and the 5-fold validation, we found a high success rate for the three methods. The resulting balanced accuracy was 92~\%,  92~\%, and 91~\% for GB, SVM and MLP, respectively. For comparison, spectro-photometric surveys typically reach accuracy of about 80~\% \citep{klimczak2022}. These relatively high classification scores are not surprising. The accuracy of classifying objects as V-types is typically very high due to the unique V-type spectra (see, e.g., \cite{penttila2021asteroid, klimczak2021predicting, moskovitz2008distribution, oszkiewicz2014selecting}). Note also that neural networks and naive Bayes methods were already trained in preparation for \gaia data by \cite{torppa2018added, penttila2021asteroid} also with very good results. Some misclassification occurs for Q/S-complex asteroids and objects with poor quality \gaia spectra. Some rare types with 0.9~\textmu{}m band such as A, R, and O-type might also be misclassified as V-types. This is natural as many of the taxonomic types start to differ more in the near-infrared (NIR) part of the spectrum which is not observed by \gaia. These rare types, however, are scarce and should not pollute our sample significantly. Furthermore, R and A-types may also be linked to differentiated parent bodies \citep{sanchez2014olivine, demeo2019olivine, moskovitz2008spectroscopically}. 

Generally all methods predicted roughly a similar number of V-types, but GB predicted the fewest and SVN highest number of V-types. By visual inspection we noticed that a small portion of the V-types predicted by SVN (especially those that have poor quality spectra) may be misclassified. This might be due to the fact that SVN divides the parameter space by hyperplanes, which are difficult to find for categories split by highly non-linear boundaries. We also compared the predicted V-types with those predicted by \cite{popescu2018taxonomic} based on NIR spectro-photometry from the MOVIS survey. We found 157 objects in common, out of which 151 were classified as V-types based Y, J, H, and Ks observations by \cite{popescu2018taxonomic} confirming the high prediction rate. Objects identified earlier by \cite{popescu2018taxonomic} based on the MOVIS survey data are marked in Table A\ref{list}. As a side note, observe that the NIR MOVIS data are highly complementary in the wavelength ranges to the \gaia DR3 VIS spectroscopy.

\begin{figure}
\includegraphics[width=0.5\textwidth]{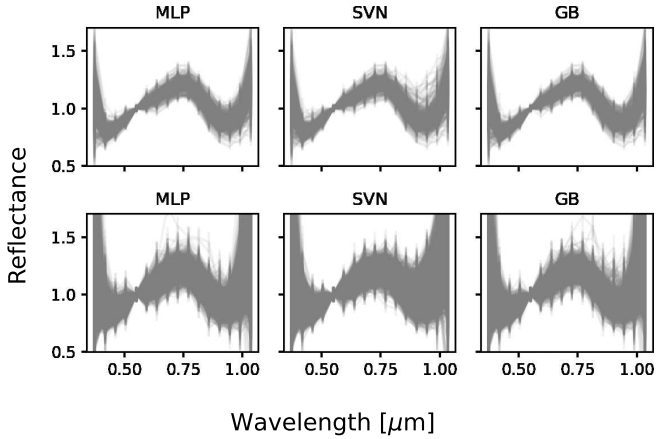}
\caption{Spectra (normalized to unity at 0.55~\textmu{}m) 
of all the classified as V-types with 
MLP (left), SVM (middle), and GB (right), both in the 
validated (top) and full sample (bottom).}
\label{spectra_gaia}
\end{figure}

In Fig. \ref{spectra_gaia} we display spectra of all the classified objects. Top panel represents the validated spectra (all quality flags validated) and in the bottom panel we show all spectra classified as V-type irrespective of quality flags. Note that for some objects reflectance at 0.55~\textmu{}m (normalization point) is flagged as invalid. Naturally, there is a larger spectral variability in the full sample.

We further split the two data sets into different populations in a consistent way with the literature \citep{ieva2016spectral, ieva2018basaltic} based on orbital parameters, semi-major axis $a$, eccentricity $e$, and inclination $i$:
\begin{itemize}
    \item Inner main belt ($2.1\,\text{au} < a \le 2.5\,\text{au}$):
    \begin{itemize}
        \item Vestoids: members of the dynamical Vesta family as defined by the hierarchical clustering method by \citet{nesvorny2012nesvorny}
        \item Fugitives: objects outside the dynamical Vesta family having $2.1\,\text{au} < a \le 2.3\,\text{au}, 5\degree < i < 8\degree$, and $0.035 < e < 0.162$
        \item Low-$i$: objects outside the dynamical Vesta family having $2.3\,\text{au} < a \le 2.5\,\text{au}$ and $i < 6\degree$
        \item Phocaea: with inclination above the $\nu_6$ resonance, $2.5\,\text{au} > a > 2.25\,\text{au}, e > 0.1$ and $32\degree > i > 18\degree$
        \item Hungaria: $2.0\,\text{au} > a > 1.78\,\text{au}, 32\degree > i > 16\degree, e < 0.18$
        \item Inner other: remaining objects in the inner main belt
    \end{itemize}
    \item Middle main belt ($2.5\,\text{au} < a \le 2.82\,\text{au}$) 
    \item Other main belt ($2.82\,\text{au} < a < 3.2\,\text{au}$)
\end{itemize}

\begin{figure}
    \centering
    \includegraphics[width=0.5 \textwidth]{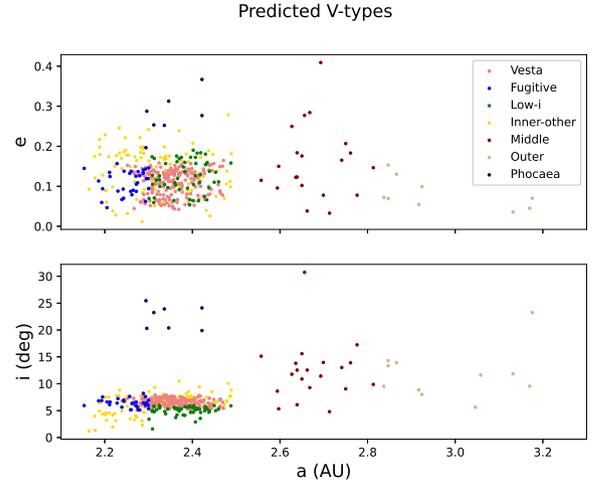}
    \caption{Distribution of orbital parameters of validated predicted V-types.}
    \label{orb_dist}
\end{figure}

  \begin{table}
    \centering
    \begin{tabular}{| l | l l l |} \hline
        Population/Method &  MLP  &  SVM  &  GB  \\ \hline
        Vesta & 170 (907) & 155 (853) & 153 (762) \\
        Fugitive & 31 (187) & 25 (157) & 31 (156) \\
        Low-$i$ & 47 (267) & 49 (293) & 46 (231) \\
        Inner-other & 105 (582) & 104 (594) & 98 (507) \\
        Middle & 16 (27) & 19 (242) & 16 (35) \\
        Outer & 10 (15) & 10 (74) & 7 (15) \\
        Hungaria & 0 (1) & 0 (6) & 0 (0) \\
        Phocaea & 5 (9) & 4 (28) & 3 (4) \\ \hline
        Total: & 384 (1995) &366 (2247) & 354 (1710) \\ \hline
    \end{tabular}
    \caption{The total number of new classified V-types in the validated and full sample (in brackets) per population.}
    \label{stats}
\end{table}

\begin{figure}
\input{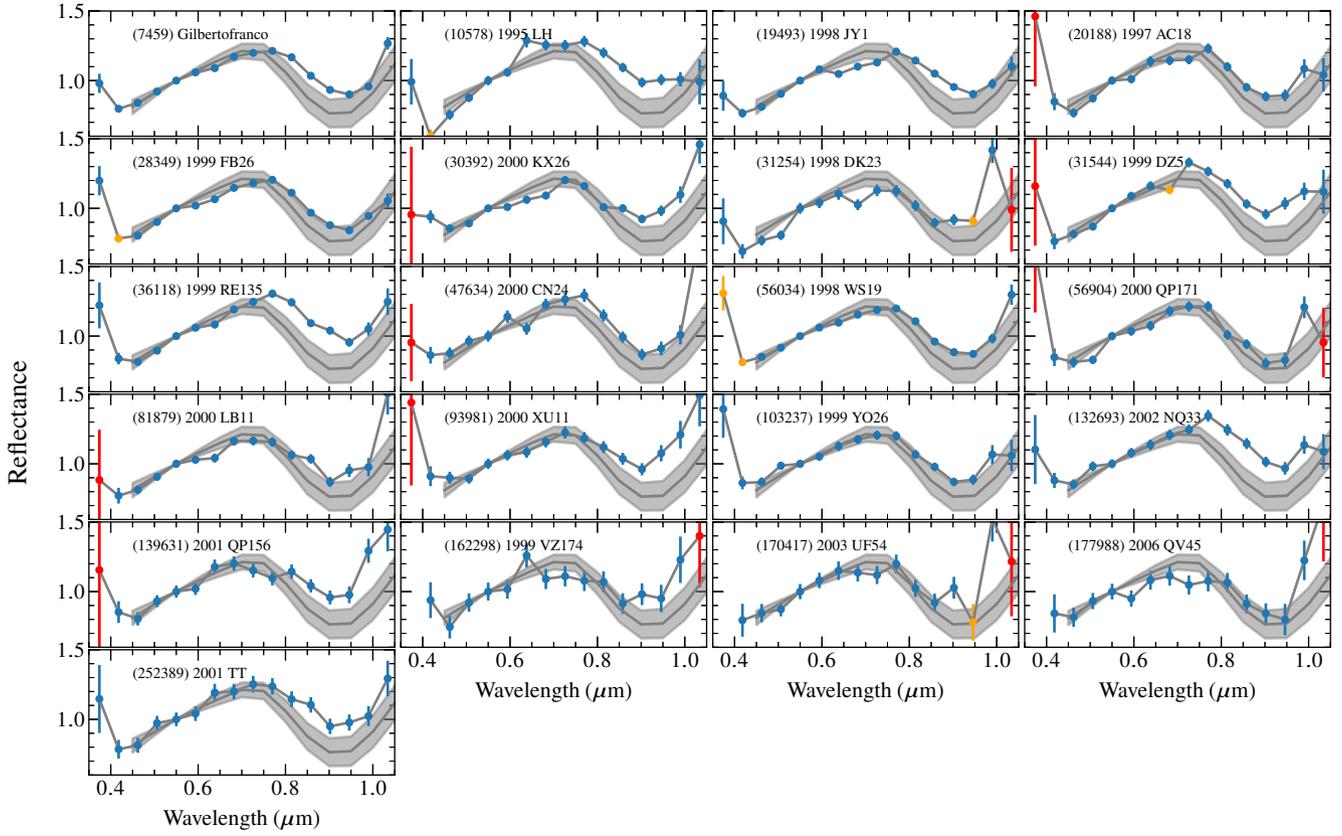}
\caption{Examples of spectra of predicted V-types in the middle belt. DR3 validated reflectances are in blue and flagged points are in orange and red (DR3 reflectrance\_spectrum\_flag equal to 1 and 2 correspondingly).}
\label{mid}
\end{figure}

\begin{figure}
 \input{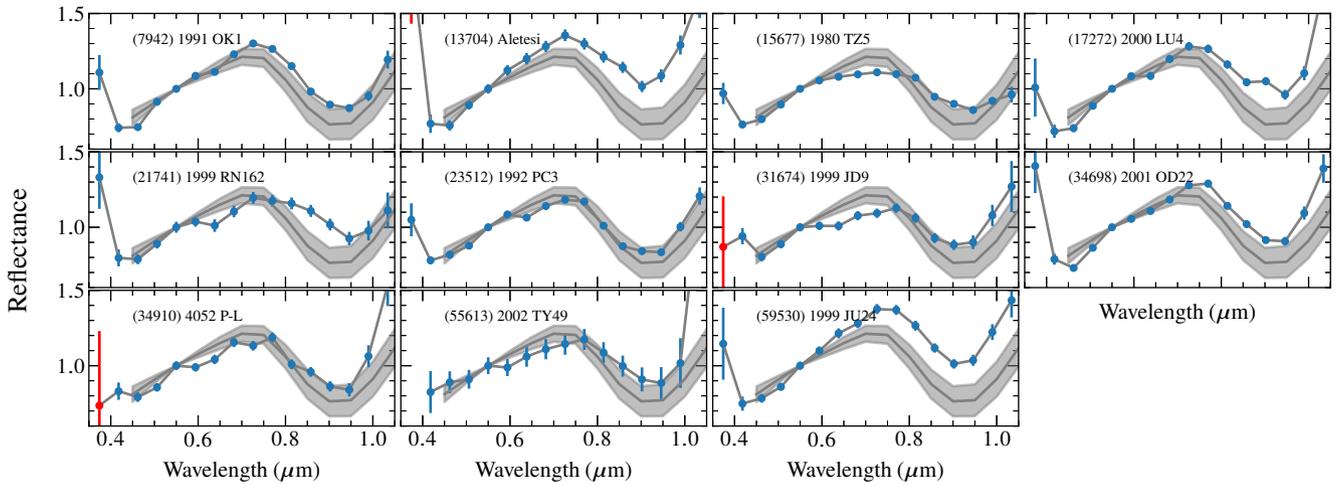}
    \caption{As in Fig. \ref{mid} but for the outer belt}
    \label{outer}
 \end{figure}

\begin{figure}
  \input{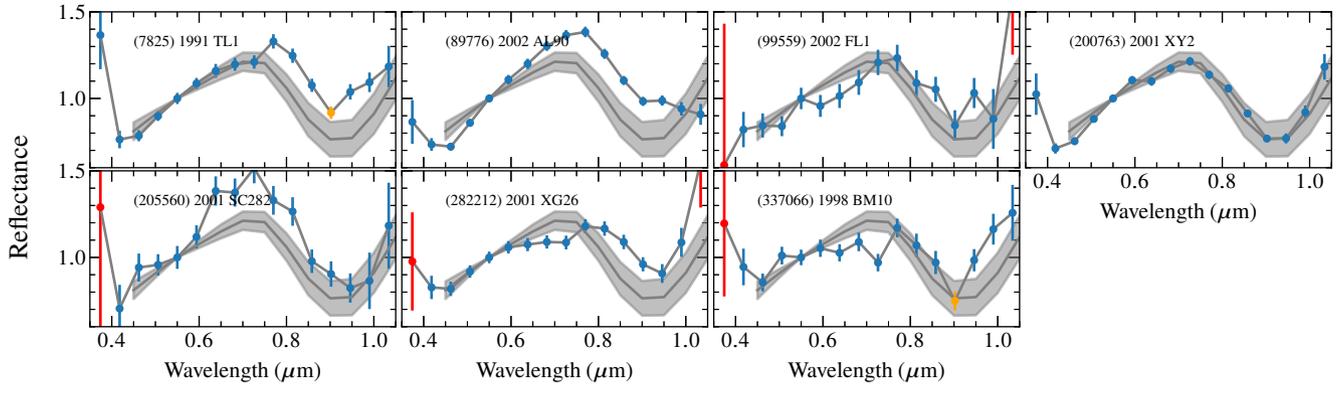}
   \caption{As in Fig. \ref{mid} but for the Phocaea region}
    \label{pha}
 \end{figure}

This division is illustrated for all validated V-types in Fig. \ref{orb_dist}. In Table~\ref{stats} we report the number of classified V-types per population in the full and validated samples. Classifications made on spectra that is not fully validated should be treated as preliminary and confirmed visually. Interestingly, in the full sample we classified a number of new V-types in the mid and outer main belt (MOVs). Since MOVs are typically considered unrelated to Vesta, and are thus key in understanding the number of differentiated planetesimals, we further inspect the predicted V-type spectra visually. We confirm 11 new V-types in the outer main belt and 20 in the mid main belt. Their spectra are plotted in Figs.~\ref{mid} and \ref{outer}. We note that asteroid (7459) Gilbertofranco was previously observed by \cite{medeiros2019compositional} and is a confirmed eucritic V-type. That object was not present in our training sample and was correctly classified as V-type by all three algorithms. Intriguingly, in the full sample we find several V-types in the Phocaea and Hungaria regions that is primarily dominated by S and X-type objects with a minor percentage of other types \citep{carvano2001spectroscopic, lucas2017hungaria, lucas2019hungaria}. Phocaea dynamical group is considered a result of a collisional breakup of a ordinary chondrite-like body \citep{carvano2001spectroscopic, carruba2009analysis}. Therefore the predicted V-types are most likely interlopers. We investigate these predicted V-types in the Phocaea region visually and find six as V-types and one as A-type (Fig.~\ref{pha}). The A-type asteroid is \nuna{89776}{ 2002 AL90} and we remove this object from further analysis. In the Hungaria region none of the predicted V-types can be confirmed.

 \onecolumn

\twocolumn

For the new V-types in the mid and outer-Main belt and in the Phocaea region we suggest further follow-up with ground-based telescopes since many of those are from the full sample with DR3 spectra that is not fully validated. In Figs. \ref{mid}-\ref{pha} we marked the points with DR3 validation flag equal to 1 and 2 (poorer quality or compromised data) in orange and red, respectively. Some objects despite validated reflectance (e.g., 205560 except for a one point in UV) show a peculiar spectra. Many show the 0.9~\textmu{}m absorption band, but do not strictly fit the ground-based templates for any taxonomic type. As discussed in Sec. \ref{spectral_params} Gaia underestimates band depths and is burdened with systematic and random offsets as compared to the ground-based spectra, thus the DR3 data will not match the ground-based templates perfectly.

\begin{table}
    \centering
    \begin{adjustbox}{width=0.5\textwidth,center}
    \begin{tabular}{lllll} \hline
    Population  & slope A & slope B & band depth & B1 center\\ 
      & (\% $10^{-3}${\AA}), & (\% $10^{-3}${\AA}), &  & (\textmu{}m) \\ \hline
    \multicolumn{4}{|c|}{MLP} \\ \hline
    Vesta       & 12.56 $\pm$ 2.46 & -26.11 $\pm$ 4.73 & 1.31 $\pm$ 0.09 & 0.93 $\pm$ 0.01 \\
    Fugitive    & 13.46 $\pm$ 2.31 & -28.91 $\pm$ 5.14 & 1.40 $\pm$ 0.11 & 0.93 $\pm$ 0.02 \\
    Low-$i$       & 13.94 $\pm$ 1.85 & -30.56 $\pm$ 5.35 & 1.43 $\pm$ 0.10 & 0.94 $\pm$ 0.01 \\
    Mid         & 12.36 $\pm$ 1.52 & -25.73 $\pm$ 1.63 & 1.31 $\pm$ 0.06 & 0.94 $\pm$ 0.01 \\
    Outer       & 16.67 $\pm$ 3.70 & -19.69 $\pm$ 6.38 & 1.41 $\pm$ 0.11 & 0.93 $\pm$ 0.01 \\
    Inner-other       & 13.76 $\pm$ 2.26 & -28.32 $\pm$ 5.15 & 1.42 $\pm$ 0.12 & 0.94 $\pm$ 0.01 \\ \hline
    \multicolumn{4}{|c|}{SVM} \\ \hline
    Vesta       & 12.67 $\pm$ 2.45 & -26.25 $\pm$ 4.65 & 1.32 $\pm$ 0.09 & 0.93 $\pm$ 0.01 \\ 
    Fugitive    & 13.52 $\pm$ 2.15 & -29.87 $\pm$ 4.80 & 1.44 $\pm$ 0.10 & 0.93 $\pm$ 0.02 \\
    Low-$i$       & 14.21 $\pm$ 1.83 & -30.56 $\pm$ 5.08 & 1.43 $\pm$ 0.10 & 0.94 $\pm$ 0.01 \\
    Mid         & 12.18 $\pm$ 2.20 & -25.19 $\pm$ 2.21 & 1.29 $\pm$ 0.01 & 0.94 $\pm$ 0.01 \\
    Outer       & 16.67 $\pm$ 4.21 & -20.34 $\pm$ 6.34 & 1.41 $\pm$ 0.10 & 0.92 $\pm$ 0.01 \\
    Inner-other       & 13.76 $\pm$ 2.21 & -28.01 $\pm$ 5.29 & 1.42 $\pm$ 0.12 & 0.94 $\pm$ 0.01 \\ \hline
    \multicolumn{4}{|c|}{GB} \\ \hline
    Vesta       & 12.76 $\pm$ 2.33 & -26.40 $\pm$ 4.74 & 1.33 $\pm$ 0.09 & 0.94 $\pm$ 0.01 \\
    Fugitive    & 13.49 $\pm$ 2.28 & -29.20 $\pm$ 5.60 & 1.40 $\pm$ 0.12 & 0.93 $\pm$ 0.01 \\
    Low-$i$       & 13.96 $\pm$ 1.83 & -30.56 $\pm$ 5.12 & 1.43 $\pm$ 0.10 & 0.94 $\pm$ 0.01 \\
    Mid         & 12.18 $\pm$ 2.20 & -25.19 $\pm$ 2.21 & 1.29 $\pm$ 0.10 & 0.94 $\pm$ 0.01 \\
    Outer       & 16.92 $\pm$ 2.31 & -22.49 $\pm$ 7.30 & 1.42 $\pm$ 0.10 & 0.93 $\pm$ 0.01 \\
    Inner-other       & 13.76 $\pm$ 2.19 & -28.41 $\pm$ 5.20 & 1.42 $\pm$ 0.12 & 0.94 $\pm$ 0.01 \\  \hline
    \end{tabular}
    \end{adjustbox}
    \caption{Median values and standard deviations for slope A, slope B, 0.9~\textmu{}m band depth and center for different populations and learning algorithms in the validated sample.}
    \label{medians}
\end{table} 

\begin{figure}
  \input{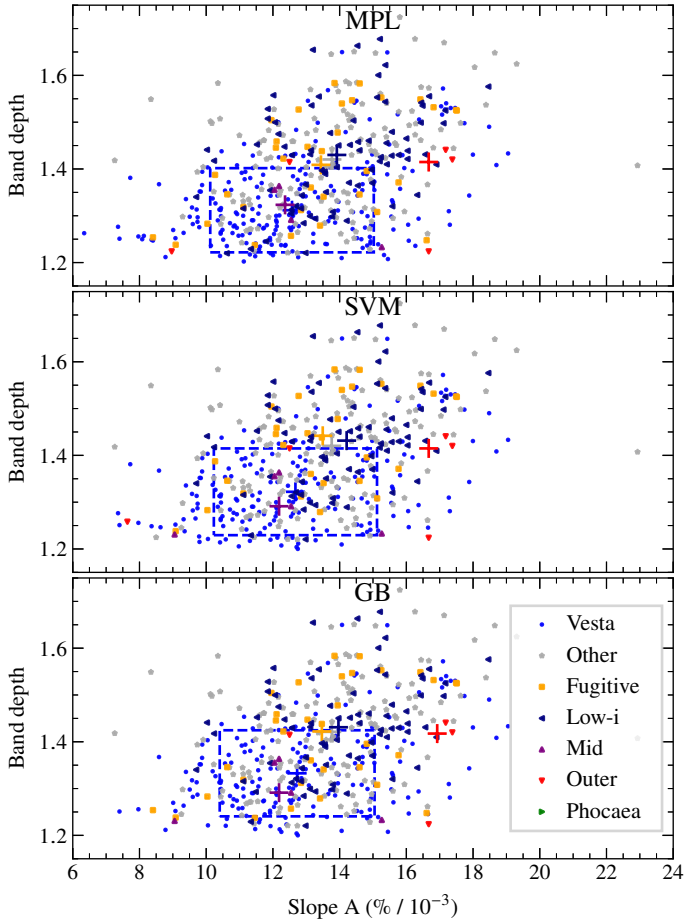}
    \caption{Slope A and apparent band depth for the V-type classified with MLP, SVM, GB methods and objects labeled in the literature. Different populations are marked with different colors and markers. Median values for the Slope A and band depth are denoted with crosses for each population. Blue frame represents $1\,\sigma$ envelope for the Vesta family members.}
    \label{spec_params1}
\end{figure}

\begin{figure}
  \input{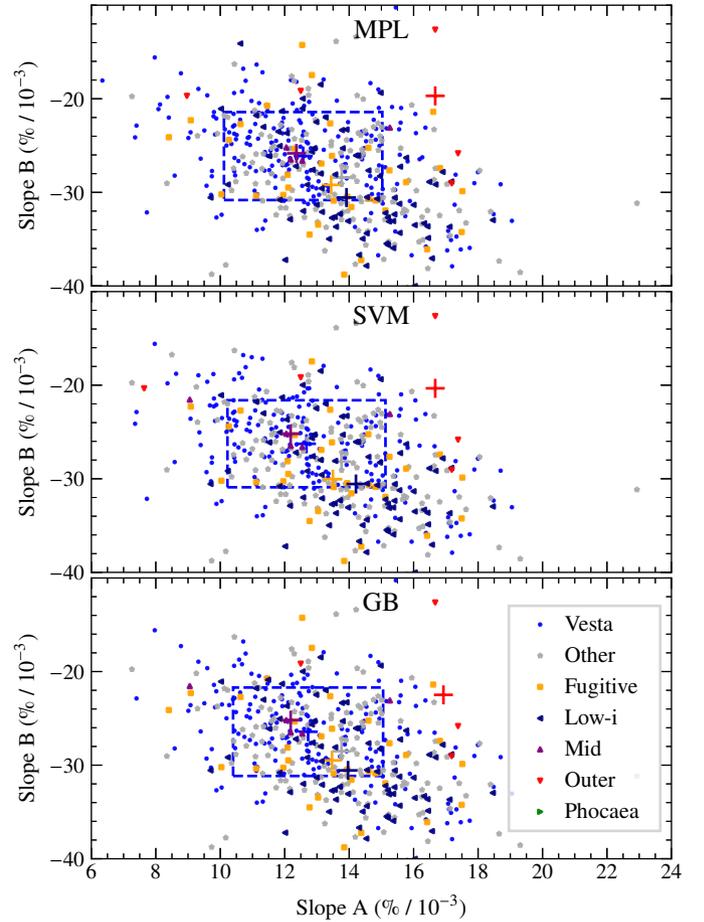}
    \caption{As in Fig. \ref{spec_params1} but for Slope A and Slope B.
    }
    \label{spec_params2}
\end{figure}

\begin{figure}
    \input{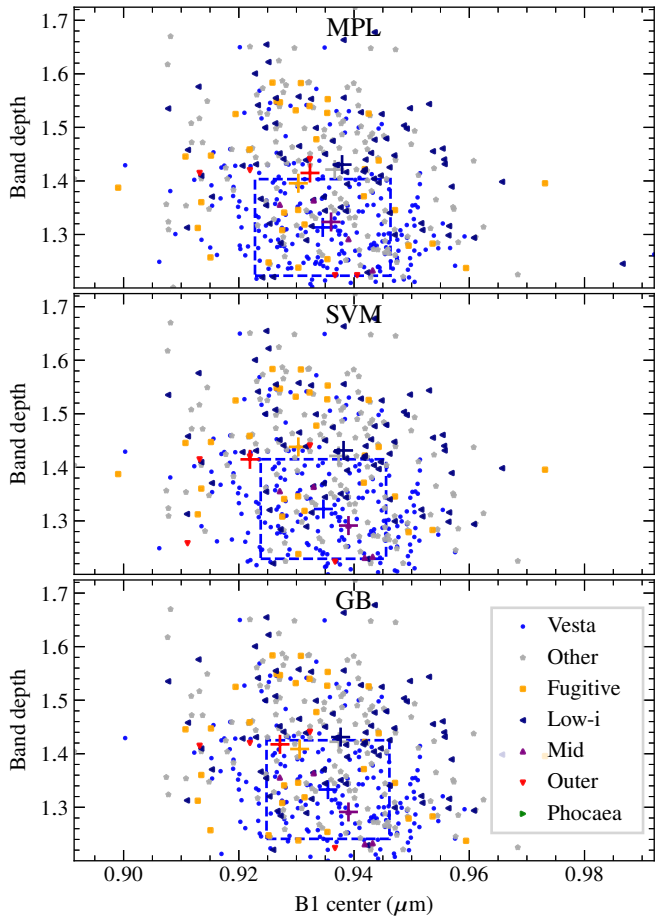}
    \caption{As in Fig. \ref{spec_params1} but for band depth and band center.}
    \label{spec_params3}
\end{figure}

We further used the classified spectra to determine spectral parameters as described in Sec.~\ref{met}. That is, we estimate slope A, slope B, apparent depth and center of the 0.9~\textmu{}m absorption band. In Figs. \ref{spec_params1} to \ref{spec_params3} we plot the estimated spectral parameters for refined sample of objects with snr $>$ 40 and quality flags validated in the wavelength ranges of the investigated parameters. We include in this sample also the V-types known from the literature with DR3 spectra fulfilling the quality criteria. We remove objects for which band depth is less than 1.2 from the analysis as those are often misclassified S-complex objects. In Table \ref{medians} we report the median values of slope A, slope B, 0.9~\textmu{}m band depth and center (B1 center) for each population and learning algorithm. Overall the median values are consistent across the different algorithms. Median values of slope A, slope B and band depth for the outer main belt population lie outside the one-standard-deviation ($1\,\sigma$) envelope of the Vesta family for all learning methods. Fugitives, low-$i$, and inner-other population have their average band depth outside or at the border of the $1\,\sigma$ envelope for the dynamical Vesta family members. Median B1 center, slope A and B for those populations are well within vestoidal values.

We find that the V-types from the outer main belt have band depths outside the $1\,\sigma$ envelope for the Vesta family members. This is consistent with the literature \citep{hardersen2004mineralogy, ieva2016spectral, leith2017compositional, ieva2018basaltic, migliorini2018spectroscopy} and has been attributed to mineralogical differences before. Additional objects in this group with large band depths strengthen the finding that those objects are distinct from the fossil planetesimal (4) Vesta and Vesta family members. While most MOVs show larger apparent depth than typical vestoids, we also find objects  (e.g. 15667 (5046 T-3) and 19493 (1998 JY1)) with a smaller band depth, more consistent with the spectra of (4) Vesta and vestoids. Additionally, one object from the middle main belt has spectral parameters well within typical vestoidal values. Two objects in the outer main belt (7942 and 34698), both at $a\sim 3.17\,\text{au}$ show similar spectral parameters. These objects are share similar semi-major axis and eccentricity, but vary in orbital inclination. Therefore a common origin would require presence of strong resonances near-by. This could be investigated with numerical studies. 

\begin{figure}
    \centering
    \includegraphics[width=0.5\textwidth]{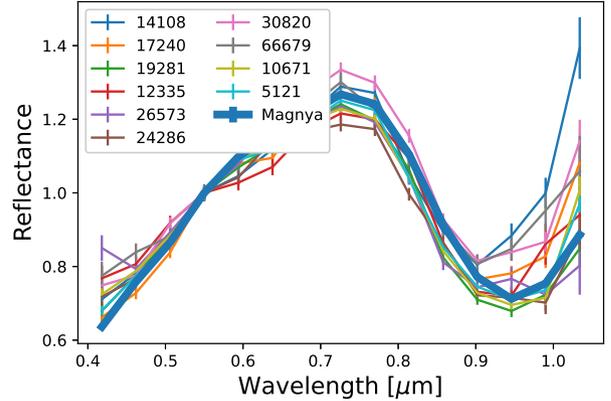}
    \caption{Spectra of selected V-types in the inner main belt with deep 0.9~\textmu{}m absorption band more consistent with that of (1459) Magnya.}
    \label{inner_selected}
\end{figure}

In the inner main belt, there is large variability of spectral parameters for all populations. Fugitives, low-$i$ and inner-other populations mostly overlap with the Vesta family objects in the spectral parameters space. Yet, a number of objects from those populations have band depths well outside the $2\,\sigma$ envelope of the Vesta family members. Furthermore, some of the objects have band depths even greater than that of (1459) Magnya. Few objects within the dynamical Vesta family also show spectral parameters well outside the $2\,\sigma$ envelope of the Vesta family. An exceptional example is asteroid 26573 with band depth more consistent with that of Magnya. We display the spectra of several such objects in Fig. \ref{inner_selected}.

Lithological differences between the vestoids and the low-$i$ asteroids were reported previously by \cite{mansour2020distribution}. Composition of one of those objects (19281) was investigated before with ground-based visible to near-infrared (VISNIR) wavelengths observations by \cite{medeiros2019compositional}. Note also that this object was not in our training sample and was correctly identified by the machine-learning algorithms as a V-type object. The ground-based spectra shows indeed large 0.9~\textmu{}m band depth. However, based on the 0.9~\textmu{}m and the 2.0~\textmu{}m band centers the authors concluded that this object is likely a former Vesta family member. One of the conclusions of \cite{medeiros2019compositional} was that compositional interpretations should rely on diagnostic plots that do not suffer from variability under slope removal. The authors recommended analysis of 0.9~\textmu{}m and 2.0~\textmu{}m band centers as the right diagnostic tool. For this reason we decided to compute also centers of the 0.9~\textmu{}m band (Fig. \ref{spec_params3}). Most asteroids from the inner main belt have their 0.9~\textmu{}m band centers well within typical vestoidal values. 

\begin{figure}
    \centering
    \includegraphics[width=0.5\textwidth]{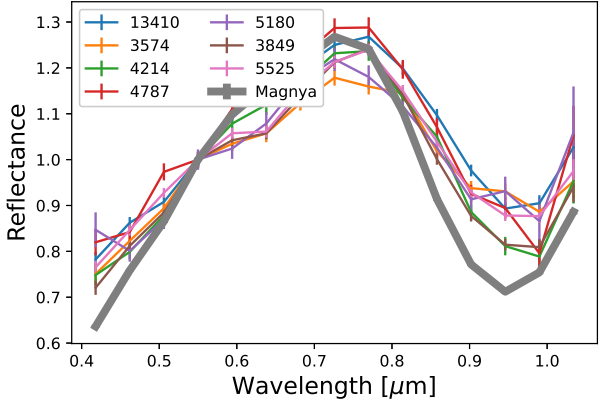}
    \caption{Spectra of selected V-types in the inner main belt with 0.9~\textmu{}m absorption band center at larger wavelengths.}
    \label{inner_B1}
\end{figure}

Objects with B1 center larger than 1.0~\textmu{}m are typically asteroids with what either appears to be erroneous DR3 spectra beyond 0.9~\textmu{}m or incorrect taxonomic classification (i.e., 89776 for which the minima is not reached by the spectrum and could be an olivine-dominated A-type asteroid). However, we also found asteroids with valid DR3 spectra having the B1 center at longer wavelengths than that of typical vestoids. DR3 spectra of some of those objects are plotted in Fig. \ref{inner_B1}. Two of those objects (5525, 3849) were previously observed \citep{oszkiewicz2020, xu1995small, bus2002phase}. For the first one, B1 center is consistent with the ground-based observations, and for the latter one, it is overestimated.

Spectral parameters for all objects are listed in Table \ref{params}.

We further comment on the spectral parameters of asteroids with large band-depths as compared to the control sample from \cite{ieva2016spectral}. The control sample in \cite{ieva2016spectral} was composed of vestoids observed at small phase angles ($1\degree<\alpha<12\degree$). The sample bounded a confined region in spectral parameters phase space. Slope A was found to vary from 7.26 and 18.51~\% $10^{-3}${\AA}, slope B from -35.98 to -18.41~\% $10^{-3}${\AA} and band depth from 1.15 to 1.72. Since band depth is underestimated in the \gaia spectra (see Sec.~\ref{data}) the upper limit for band depth for the \gaia spectra should be about 0.2 less than that from \cite{ieva2016spectral}. This leaves a number of V-types from the inner main belt outside the region of spectral parameters bounded by the control sample from \cite{ieva2016spectral}. 

We conclude that indeed the inner main belt could be a mix of objects from various planetesimals, not just (4) Vesta. Number of identified asteroids with large 0.9~\textmu{}m band depths or band centers offset from typical vestoids is however low. The differentiated planetesimals thus could have formed in a wider orbital range (e.g., 1.3~au and 7.5~au from the Sun as indicated by \citep{lichtenberg2021bifurcation}) as opposed to a narrow terrestrial planet region proposed by \cite{bottke2006iron} which would lead to increased number of those objects in the inner main belt. It is also plausible that these objects remain spectrally indistinguishable from Vesta and Vesta family members. 

These objects should be followed-up by ground-based VISNIR spectroscopy at small phase angles for full mineralogical characterization.

\section{Discussion}
\label{disc}

The largest source of uncertainty in this work is the phase reddening effect (changes in spectral slope, color indices, variations in albedo, and absorption bands depth with changing phase angle). Although known for decades, the effect has been studied in detail on a very limited number (less than several dozens) of objects \citep{millis1976ubv, miner1976five, clark2002near, reddy2012photometric, sanchez2012phase} and for only a few taxonomic types. Moreover, the effect has been suggested to differ from asteroid to asteroid by \cite{BINZEL201941, carvano2015shape}. A limited laboratory work also demonstrated the effect on various analogues of planetary material \citep{gradie1986wavelength, gradie1980effects}.

Since average phase angles are not reported with the \gaia spectra, correcting for this effect is not possible. Moreover, a single \gaia spectra is obtained from averaging spectra from multiple transits, thus from multiple phase angles, therefore correcting for phase reddening becomes nontrivial. 

Median phase angle for an average \gaia spectra is about 20 degrees (whereas the ground-based observations are typically taken at opposition, close to 0 degree) thus there is a concern of phase reddening affecting the obtained asteroid spectra \citep{gaiaspec, cellino2020ground}. At these phase angles phase reddening is expected to increase spectral slopes and band depths of V-type asteroids \citep{reddy2012photometric}. 

Based on satellite observations of a single object (4) Vesta, \cite{reddy2012photometric} found around 20~\% increase in overall spectral slope for every 10 degrees increase of phase angle and 2.35~\% increase of the band depth of 0.9~\textmu{}m absorption band. Since band depth are not influenced significantly, our conclusions for inner main belt V-types with large absorption bands would still hold. Furthermore we found that band depths are systematically underestimated in the \gaia spectra, thus they may be even larger than estimated.

\cite{ieva2016spectral, ieva2018basaltic} used the relation from \cite{reddy2012photometric} to correct the spectral parameters of V-types for phase reddening, but also noted that the correction is more qualitative than quantitative. Since band depths are not significantly affected, the same conclusions would have been reached without the correction.

Other effects such as space weathering and regardening have less impact on our conclusions. Space weathering is expected to lower band-depths \citep{vernazza2006asteroid, fulvio2012space} and regardening is acting on near-Earth asteroids \citep{binzel2010earth}.

Another source of concern are the unknown systematic and random errors of the \gaia spectra. The \gaia mission was designed for stellar tracking. Spectra for moving objects such as asteroids is extracted from windows that are moving at stellar rates. Since the XP spectra is taken at the end of the transit the difference between the actual asteroid position and the position of the window is the largest \citep{gaiaspec}. Complicated wavelength calibration can cause further uncertainties \citep{gaiaspec}. As noted in Sec.~\ref{data} DR3 spectra tends to underestimate band depth for V-type asteroids. \cite{gaiaspec} estimated the average difference between SMASS II and DR3 z-i color across all taxonomic types to be about $-0.08$ and derived formulas for correcting the z-i color and overall spectral slope to the SMASS II dataset.

These formulas are however more applicable to population studies and not to individual objects, due to large variability of the correction. Furthermore, the V-type asteroids seem to be affected more by this systematic effect than other taxonomic types \citep{gaiaspec}. 

The ground-based spectral observations over the visible and near-infrared wavelengths will allow a detailed characterization of the most interesting targets reported in this work, namely asteroids from Figs. \ref{inner_selected} and \ref{inner_B1} in the inner main belt as well as all potential V-types in the mid and outer main belt reported in Tables \ref{list}. For observers convenience we list the dates of nearest oppositions, predicted V-magnitude, and declination at opposition for the interesting objects in Tables \ref{obs-in} to \ref{obs-pho}.

\begin{table}
    \centering
    \begin{tabular}{lllll} \hline
        Number & designation & next opp. & decl. (deg) & V (mag) \\ \hline
         13410  & 1999 UX5 Arhale       & 2022/12/19  & +22  & 17.4 \\ 
         3574   & 1982 TQ  Rudaux       & 2023/08/03  & -09  & 16.0 \\
         4214   & 1987 UX4 Veralynn     & 2023/05/02  & -18  & 16.2 \\
         4787   & 1986 RC7 Shul'zhenko  & 2023/07/17  & -30  & 16.1 \\
         5180   & 1989 GF  Ohno         & 2023/07/18  & -27  & 16.8 \\
         3849   & 1984 FC  Incidentia   & 2023/04/12  & -08  & 16.0 \\
         5525   & 1991 TS4              & 2023/03/16  & -01  & 16.4 \\
         14108  & 1998 OA13             & 2022/09/09  & +01  & 16.3 \\
         172240 & 2002 RO19             & 2023/05/23  & -15  & 20.2 \\
         19281  & 1996 AP3              & 2022/08/04  & -08  & 15.3 \\
         12335  & 1992 WJ3 Tatsukushi   & 2023/09/04  & -05  & 16.9 \\
         26573  & 2000 EG87             & 2022/09/20  & +03  & 15.8 \\
         24286  & 1999 XU18             & 2023/08/26  & -12  & 16.0 \\
         30820  & 1990 RU2              & 2022/08/19  & -19  & 16.7 \\
         66679  & 1999 TD29             & 2022/10/27  & +10  & 17.4 \\
         10671  & 1977 RR6 Mazurova     & 2023/12/23  & +32  & 16.8 \\
         5121   & 1989 AX1 Numazawa     & 2023/04/10  & -11  & 16.5 \\
         19281  & 1996 AP3              & 2022/08/04  & -08  & 15.3 \\ \hline
    \end{tabular}
    \caption{Future observing opportunities for selected inner-main Belt V-types. Columns are: asteroid number, designation, date of the next opposition, declination and predicted V-magnitude at the nearest opposition.}
    \label{obs-in}
\end{table}

\begin{table}
    \centering
    \begin{tabular}{lllll} \hline
        Number & designation & next opp. & decl. (deg) & V (mag) \\ \hline
    10578   & 1995 LH   & 2022/11/30  & +28  & 18.0 \\
    28349   & 1999 FB26 & 2023/05/07  & -29  & 17.9 \\
    81879   & 2000 LB11 & 2023/06/02  & -08  & 18.3 \\ 
    36118   & 1999 RE13 & 2022/11/06  & +09  & 17.4 \\
    170417  & 2003 UF54 & 2023/01/23  & +03  & 19.6 \\
    139631  & 2001 QP15 & 2023/05/27  & -01  & 20.1 \\
    30392   & 2000 KX26 & 2023/10/02  & +21  & 18.1 \\
    252389  & 2001 TT   & 2022/10/05  & -35  & 19.6 \\
    93981   & 2000 XU11 & 2023/01/06  & +37  & 17.9 \\
    19493   & 1998 JY1  & 2023/03/14  & -04  & 17.7 \\
    47634   & 2000 CN24 & 2023/06/25  & -06  & 18.4  \\
    56904   & 2000 QP17 & 2022/11/11  & +01  & 18.9  \\
    7459    & 1984 HR1 Gilbertofranco & 2023/10/29  & +13  & 16.8  \\
    20188   & 1997 AC18 & 2023/06/25  & -21  & 17.0  \\
    132693  & 2002 NQ33 & 2023/04/28  & -29  & 18.8  \\
    177988  & 2006 QV45 & 2023/09/01  & -05  & 19.3 \\
    103237  & 1999 YO26 & 2023/10/25  & +20  & 17.2 \\ \hline
    \end{tabular}
    \caption{As in Table \ref{obs-in} but for objects in the mid main belt.}
    \label{oubs-mid}
\end{table}

\begin{table}
    \centering
    \begin{tabular}{lllll} \hline
        Number & designation & next opp. & decl. (deg) & V (mag) \\ \hline
    7942    & 1991 OK1  & 2023/02/13  & +09  & 16.4 \\
    15677   & 1980 TZ5  & 2023/09/10  & -18  & 15.6 \\
    23512   & 1992 PC3  & 2023/05/07  & -29  & 17.1  \\
    34910   & 4052 P-L  & 2023/04/08  & -10  & 18.7  \\
    55613   & 2002 TY49 & 2022/11/21  & +20  & 18.3 \\
    59530   & 1999 JU24 & 2023/09/28  & -06  & 17.5  \\
    23512   & 1992 PC3  & 2023/05/07  & -29  & 17.1   \\
    13704   & 1998 PA1 Aletesi & 2023/05/01  & -17  & 18.5  \\
    21741   & 1999 RN16 & 2023/09/10  & -23  & 17.2 \\
    17272   & 2000 LU4  & 2023/03/27  & +08  & 17.7 \\ \hline
    \end{tabular}
    \caption{As in Table \ref{obs-in} but for objects in the outer main belt.}
    \label{obs-out}
\end{table}

\begin{table}
    \centering
    \begin{tabular}{lllll} \hline
    Number & designation & next opp. & decl. (deg) & V (mag) \\ \hline
    7825    & 1991 TL1  & 2023/04/21  & -04  & 17.2 \\
    89776   & 2002 AL90 & 2022/09/07  & -18  & 17.4 \\
    99559   & 2002 FL1  & 2023/03/25  & +02  & 16.8 \\
    205560  & 2001 SC28 & 2023/01/03  & -11  & 17.7 \\
    282212  & 2001 XG26 & 2022/11/04  & -33  & 17.9 \\
    337066  & 1998 BM10 & 2022/09/19  & -29  & 21.2 \\ \hline
    \end{tabular}
    \caption{As in Table \ref{obs-in} but for objects in the Phocaea region.}
    \label{obs-pho}
\end{table}

\section{Conclusions}
\label{conl}

We classified 60,518 asteroids from the \gaia DR3 catalogue into V-types and non-V-types. We found about 2000 new V-types and validated around 350 of them. We found 31 new V-types in the mid and outer main belt and 6 in the Phocaea region. This adds to the inventory of objects that are not easily linked to the fossil planetesimal (4) Vesta. We suggest further follow up of those objects by ground-based observations to fully confirm their basaltic nature. Large number of V-types is found in the inner main belt.

For the first time, we found a number of objects in the inner main belt with large band depths, some even larger than that of (1459) Magnya. We also found objects with 0.9~\textmu{}m band centers offset from typical vestoidal values. However, band centers are randomly overestimated in the DR3 data thus are not reliable. Future asteroid spectroscopy to be provided in the \mbox{\gaia} DR4 catalog may contain improved processing with spectra having less biases thus might also provide a better insight into the spectroscopic parameters of V-types.

We show that there are V-type asteroids in the inner main belt that may be distinct from Vesta. Further ground-based observations in the near-infrared are needed for full mineralogical analysis of those objects. Determination of spins and shapes accompanied with backward dynamical analysis may further provide evidence for existence of differentiated planetesimals in the inner main belt other than Vesta.

\section*{Acknowledgments}

This work has been supported by grant No. 2017/26/D/ST9/00240 from the National Science Centre, Poland. HK was supported by grant No. 2017/25/B/ST9/00740 from the National Science Centre, Poland, and AP was supported by grant No. 325805 from the Academy of Finland. The work of MP was made in the framework of the project PN19-030102-INCDFM, contract number 21N/2019, MCID, Romania. We used data from the European Space Agency mission \gaia\footnote{\url{https://www.cosmos.esa.int/gaia}}

\section*{Data Availability}

This research used publicly available data from the DR3 catalog of the  European Space Agency mission \gaia. Tables \ref{list} and \ref{params} will be provided in a numerical form though CDS after publication.



\bibliographystyle{mnras}
\bibliography{biblio}



\appendix

\onecolumn

\begin{longtable}{lrrrrrrrrr}\hline
                           Full name &     $a$ (au) &     $e$ &      $i$ (deg)  & MLP & SVM & GB & MP18 & valid & snr \\
                           \hline
                           
                \multicolumn{8}{c}{Outer main belt} \\ \hline
                   14390 (1990 QP10) &  3.246 &  0.102 &   6.56 &     \checkmark &     \checkmark &         &              &      0 &   36.28 \\
                   71524 (2000 CM78) &  3.242 &  0.156 &   1.86 &         &     \checkmark &         &              &      0 &   14.56 \\
                  92169 (1999 XN181) &  3.239 &  0.031 &  20.75 &         &     \checkmark &         &              &      0 &   18.17 \\
                   113371 (2002 SC6) &  3.225 &  0.122 &   5.94 &         &     \checkmark &         &              &      0 &   14.01 \\
                   28882 (2000 KO48) &  3.224 &  0.122 &  14.51 &         &     \checkmark &         &              &      0 &   16.26 \\
                   85632 (1998 KV52) &  3.198 &  0.107 &  21.46 &         &     \checkmark &         &              &      0 &   13.64 \\
\end{longtable}

\begin{longtable}{llllrrrrl}
\hline
                         Full name &      $a$ &       $e$ &      $i$  &   slope A  &   slope B   &  band depth & B1 center& population\\
                         & (AU) & & (deg) & (\% 10$^{-3}$\AA) & (\% 10$^{-3}$\AA) &   &(\textmu{}m) & \\                         
\hline
                 34698 (2001 OD22) &  3.176 &  0.070 &  23.24 &   17.379 &  -25.822 &       1.420 &      0.922 &     outer \\
                   7942 (1991 OK1) &  3.170 &  0.045 &   9.55 &   17.180 &  -29.034 &       1.441 &      0.932 &     outer \\
             1459 Magnya (1937 VA) &  3.149 &  0.229 &  16.90 &   17.616 &  -37.877 &       1.633 &      0.949 &     outer \\
                  23512 (1992 PC3) &  3.058 &  0.152 &  11.61 &   12.498 &  -19.156 &       1.415 &      0.913 &     outer \\
                  15677 (1980 TZ5) &  2.847 &  0.153 &  13.34 &    8.966 &  -19.689 &       1.224 &      0.941 &     outer \\
                  17272 (2000 LU4) &  2.837 &  0.072 &   9.53 &   16.670 &  -12.615 &       1.224 &      0.937 &     outer \\
\end{longtable}


\bsp	
\label{lastpage}
\end{document}